\documentclass[aps,prl,reprint,superscriptaddress,nobibnotes]{revtex4-2}
\usepackage{mathrsfs}
\usepackage{amsmath,gensymb}
\usepackage{amsfonts}
\usepackage{amssymb}
\usepackage{amsthm}
\usepackage{graphicx}
\usepackage{natbib}
\usepackage{xcolor}
\usepackage{hyperref}
\usepackage{bm}
\usepackage[caption=false]{subfig}
\usepackage{verbatim}
\usepackage{siunitx}

\newcommand{\ui}{{\rm i}}

\newcounter{numsection}
\newcommand{\numsection}[1]{
    \stepcounter{numsection}
    \noindent\textbf{\large Supplemental Note \arabic{numsection}: #1}
}

\begin{document}
\title{Spin pumping reveals vortex-mediated angular-momentum dissipation at the superconducting transition}

\author{Maximilian Mangold}
\affiliation{School of Natural Sciences, Technical University of Munich, 85748 Garching b. Munich, Germany}
\affiliation{Center for Quantum Engineering (ZQE), Technical University of Munich, 85748 Garching b. Munich, Germany}
\author{Alex Burg}
\affiliation{School of Natural Sciences, Technical University of Munich, 85748 Garching b. Munich, Germany}
\affiliation{Center for Quantum Engineering (ZQE), Technical University of Munich, 85748 Garching b. Munich, Germany}
\author{Franz Weidenhiller}
\affiliation{School of Natural Sciences, Technical University of Munich, 85748 Garching b. Munich, Germany}
\affiliation{Center for Quantum Engineering (ZQE), Technical University of Munich, 85748 Garching b. Munich, Germany}
\author{Thomas N.G. Meier}
\affiliation{School of Natural Sciences, Technical University of Munich, 85748 Garching b. Munich, Germany}
\affiliation{Center for Quantum Engineering (ZQE), Technical University of Munich, 85748 Garching b. Munich, Germany}
\author{Hiroto Adachi}
\affiliation{Research Institute for Interdisciplinary Science, Okayama University, Okayama 700-8530, Japan}
\author{Lin Chen}
\email[Corresponding Author: ]{lin0.chen@tum.de}
\affiliation{School of Natural Sciences, Technical University of Munich, 85748 Garching b. Munich, Germany}
\affiliation{Center for Quantum Engineering (ZQE), Technical University of Munich, 85748 Garching b. Munich, Germany}
\author{Christian H. Back}
\affiliation{School of Natural Sciences, Technical University of Munich, 85748 Garching b. Munich, Germany}
\affiliation{Center for Quantum Engineering (ZQE), Technical University of Munich, 85748 Garching b. Munich, Germany}

\begin{abstract}
Spin pumping into superconductors is generally expected to become less efficient below the critical temperature $T_\textrm{c}$ because quasiparticle transport freezes out. Here we report a pronounced enhancement of the spin pumping-induced magnetic damping in Py/Nb heterostructures, confined to a narrow temperature interval immediately below the superconducting transition. The damping exceeds its normal-state value and therefore cannot be explained within the conventional quasiparticle picture of spin transport. The enhancement coincides with the finite-resistance vortex-creep regime of the Nb layers and disappears upon cooling into the pinned vortex state, indicating that mobile Abrikosov vortices provide an additional channel for spin-angular-momentum dissipation. These results establish ferromagnetic resonance as a sensitive probe of dissipative vortex dynamics in superconductors.
\end{abstract}

\maketitle

The transfer of spin angular momentum into superconducting materials provides a powerful probe of the elementary excitations responsible for spin transport and dissipation. In ferromagnet (FM)/superconductor (SC) heterostructures, a coherently precessing magnetization injects a pure spin current into the superconducting layer by spin pumping, allowing angular momentum transport to be investigated without accompanying charge transport\,\cite{Tserkovnyak2002,Tserkovnyak2005}. Spin pumping has consequently emerged as a versatile experimental tool for studying superconducting spin transport and magnetization dynamics in hybrid superconducting structures\,\cite{Houzet2008,Yokoyama2009,Bell2008,Jeon2018,Jeon2018b,Yao2018,Golovchanskiy2020}.

For conventional spin-singlet SCs, spin transport is generally attributed to thermally excited quasiparticles. Accordingly, the opening of the superconducting gap below the critical temperature $T_\textrm{c}$ suppresses the quasiparticle population and leads to the characteristic reduction of spin pumping-induced damping observed in Nb-based heterostructures and related systems\,\cite{Bell2008,Jeon2018,Jeon2018b,Yao2018,Golovchanskiy2020,Mueller2021}. More recently, theoretical work has predicted that superconducting coherence effects, finite-frequency spin susceptibility, Andreev resonances and surface bound states can enhance spin absorption under suitable conditions\,\cite{Inoue2017,Silaev2020a,Silaev2020b,Tanhayi2021}.
Despite these refinements, the transfer of angular momentum is still assumed to occur exclusively through electronic excitations of the superconducting condensate.

Type-II SCs, however, possess another collective degree of freedom that is intrinsically dissipative: Abrikosov vortices\,\cite{Shubnikov1937,Abrikosov1957}. In the mixed state, the motion of vortices governs electrical dissipation through flux-flow transport, while increasing pinning forces progressively suppress the vortex motion upon further cooling\,\cite{Eley2021}. The possibility that mobile vortices can also couple directly to spin currents has recently attracted considerable theoretical interest through the prediction of the vortex spin Hall effect and related phenomena\,\cite{Taira2021,Adachi2024,Kim2018}. This raises a fundamental question: Can a pure spin current transfer angular momentum directly to a mobile vortex ensemble, thereby opening an additional channel for spin-angular-momentum dissipation beyond conventional quasiparticle transport? To date, experimental evidence for such a dissipation channel has remained elusive.

\begin{figure}[t]
    \centering
	\includegraphics[width=3.4in]{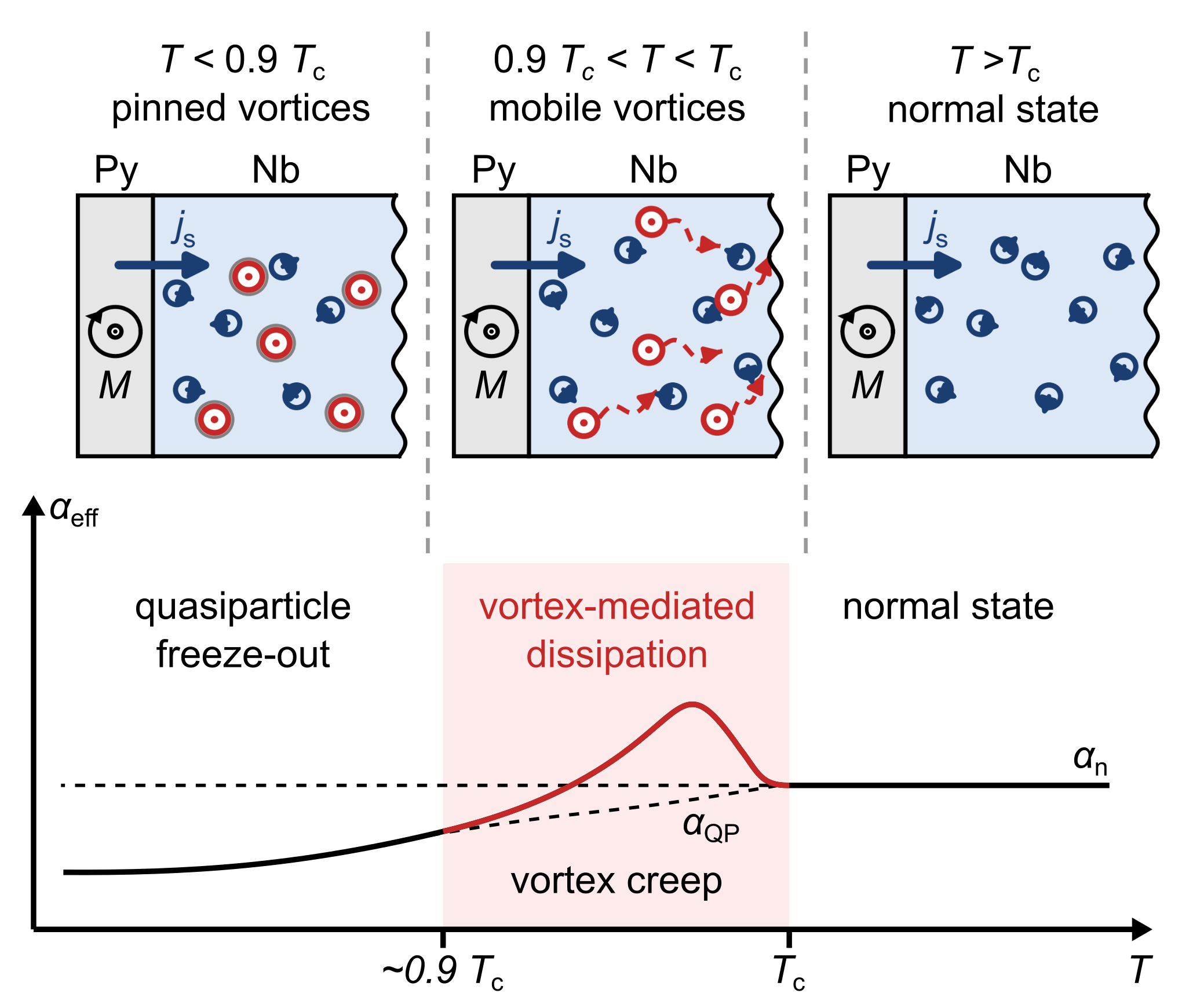}
    \caption {Conceptual picture of vortex-mediated spin-angular-momentum dissipation. Top: the precessing magnetization $M$ of the Py layer pumps a spin current $j_\textrm{s}$ into the adjacent superconducting Nb layer in three temperature regimes. Deep in the superconducting state ($T < 0.9\,T_\textrm{c}$), vortices (red) are pinned and spin angular momentum is absorbed only by thermally excited spin-carrying quasiparticles (blue), which  progressively freeze out upon cooling. In the vortex-creep regime, shown here for $0.9\,T_\textrm{c} < T < T_\textrm{c}$, mobile vortices provide an additional dissipation channel on top of the quasiparticle channel, enhancing the effective damping. In the normal state ($T > T_\textrm{c}$), vortices are absent and the damping is governed by the quasiparticle channel alone. Bottom: schematic temperature dependence of the effective damping $\alpha_\textrm{eff}$ resulting from the two channels. The excess damping above the normal-state value $\alpha_\textrm{n}$ (dashed line) is confined to the dissipative vortex-creep interval (shaded). At lower temperatures, $\alpha_\textrm{eff}$ is determined by the quasiparticle contribution $\alpha_\textrm{QP}$.}
    \label{Concept}
\end{figure}

Here we address this question using broadband ferromagnetic resonance (FMR) measurements on symmetric SC/FM/SC heterostructures. We observe a distinct enhancement of the effective Gilbert damping in a narrow temperature interval below $T_\textrm{c}$, where the damping increases above its normal-state level despite the expected suppression of quasiparticle-mediated spin transport.
Through electrical transport experiments we confirm that the enhancement coincides with the dissipative vortex-creep regime and disappears as vortex motion becomes suppressed by pinning effects at lower temperatures. This close correspondence strongly suggests that mobile vortices provide an additional channel for spin-angular-momentum dissipation, establishing spin pumping as a sensitive probe of vortex dynamics in type-II SCs. The proposed physical picture is summarized schematically in Fig.\,\ref{Concept}.

\begin{figure}[t]
	\centering
	\includegraphics[width=3.4in]{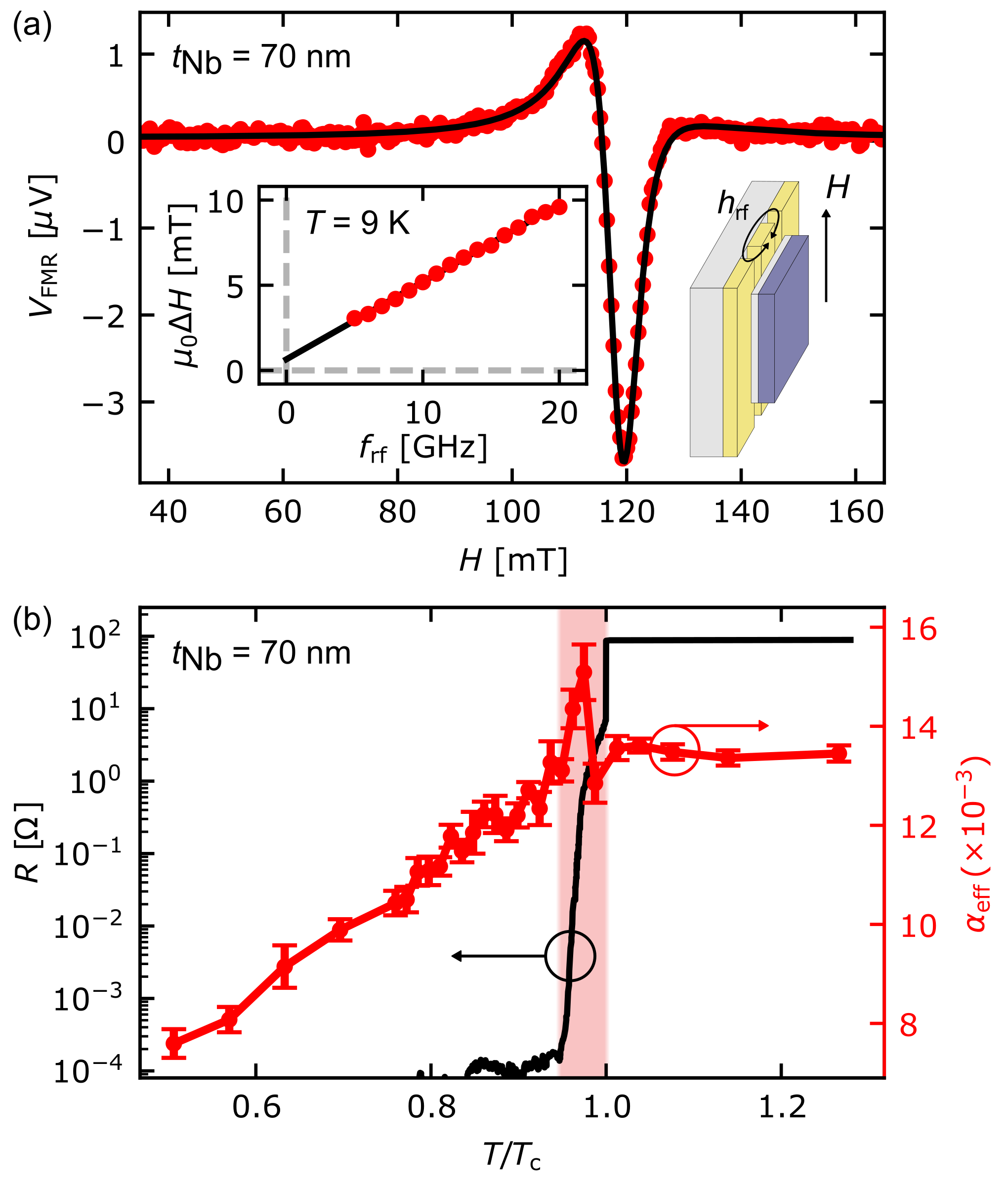}
	\caption{Enhancement of effective damping near the superconducting transition. (a) Exemplary resonance curve, acquired on the sample with $t_\textrm{Nb}=\SI{70}{nm}$ at $T=\SI{9}{K}$ and $f_\textrm{rf}=\SI{10}{GHz}$. The black line represents the Lorentzian fit, see Supplemental Material\,\cite{SupplMat}. The inset depicts the frequency-dependence of the linewidth $\Delta H$, shown for $T=\SI{9}{K}$, and linear fit function from which $\alpha_\textrm{eff}$ is extracted. The measurement geometry is sketched on the bottom right. (b) Temperature dependence of the four-point resistance (left axis) measured at $H=\SI{500}{mT}$ and effective Gilbert damping (right axis) for the sample with $t_\textrm{Nb}=\SI{70}{nm}$. The red shaded area highlights the temperature interval $[0.95\,T_\textrm{c},\,T_\textrm{c}]$ where the resistance features a low-resistance tail. Note that the transport- and FMR-derived transition temperatures differ in absolute temperature (see Supplemental Material\,\cite{SupplMat}); here, each quantity is shown versus its respective normalized temperature $T/T_\textrm{c}$.}
	\label{FMR-alpha}
\end{figure}

To test this conceptual mechanism, we investigate FMR spin pumping across the superconducting transition in Si/SiO$_2$/Ta(4)/Pt(4)/Nb($t_\textrm{Nb}$)/Py(6)/Nb($t_\textrm{Nb}$)/Pt(4) thin-film heterostructures with systematically varied Nb thickness $t_\textrm{Nb}$. This symmetric sample design minimizes spin-backflow asymmetries and isolates the contribution of the Nb to the effective Gilbert damping. Varying $t_\textrm{Nb}$ between 30 and 70\,nm allows us to systematically tune the superconducting properties and the extent of the dissipative vortex regime while maintaining identical FM layers. A schematic sketch of the broadband FMR experiment is drawn in the inset of Fig.\,\ref{FMR-alpha}(a). A microwave current at the frequency $f_\textrm{rf}$ flowing through the coplanar waveguide produces the time-varying driving field $h_\textrm{rf}$ which excites magnetization precession in the Py layer, injecting spin angular momentum into the adjacent superconducting Nb layers by spin pumping. The absorbed power $V_\textrm{FMR}$ is measured by lock-in detection. Figure\,\ref{FMR-alpha}(a) depicts an exemplary resonance curves of the sample with $t_\textrm{Nb}=\SI{70}{nm}$ acquired by sweeping the applied magnetic field through the resonance condition. The resonance field $H_\textrm{r}$ and linewidth $\Delta H$ are extracted from a Lorentzian fit. For more details on the measurement and evaluation method, see Supplemental Material\,\cite{SupplMat}.

An example of the $f_\textrm{rf}$-dependence of $\Delta H$ is shown in the inset of Fig.\,\ref{FMR-alpha}(a). The effective Gilbert damping parameter $\alpha_\textrm{eff}$ is extracted using the standard linear relation
$\mu_0\Delta H = \mu_0\Delta H_0 + \frac{2\pi\alpha_{\textrm{eff}}}{\gamma}f_\textrm{rf}$,
where $\gamma$ is the gyromagnetic ratio and $\Delta H_0$ the frequency-independent inhomogeneous linewidth broadening. To investigate the spin-dissipation channels in the SC, the frequency-dependent FMR measurements are performed at various temperatures near the superconducting transition of the Nb layers. When cooling from the normal regime at \SI{9}{K} into the superconducting state, a shift of $H_\textrm{r}$ is observed (see Supplemental Material\,\cite{SupplMat}), consistent with Refs.\,\cite{Li2018, Jeon2019, Golovchanskiy2020, Tian2025}, from which the critical temperature $T_\textrm{c} = 7.9$ (7.2, 5.5) K is extracted for $t_\textrm{Nb} = 70$ (45, 30) nm.

Figure\,\ref{FMR-alpha}(b) depicts the temperature dependence of the effective Gilbert damping for the sample with $t_\textrm{Nb} = \SI{70}{nm}$. Above $T_\textrm{c}$, the damping remains essentially constant at the normal-state level $\alpha_\textrm{n} = 13.5\times10^{-3}$.
Upon cooling through the superconducting transition, $\alpha_\textrm{eff}$ first increases above $\alpha_\textrm{n}$ before decreasing again at lower temperatures.
Besides $\alpha_\textrm{eff}$, the film resistance $R$ is shown in Fig.\,\ref{FMR-alpha}(b). A magnetic field of $H=\SI{500}{mT}$ is applied which is comparable to the upper limit of the magnetic fields applied during FMR measurements. For more details on the transport measurements, see Supplemental Material\,\cite{SupplMat}. Below the sharp superconducting transition at $T_\textrm{c}$, a finite-resistance tail is observed, indicative of the vortex-creep regime\,\cite{Eley2021, Villegas2005, Altanany2024}. 
This dissipative response is confined to the temperature interval between $T_\textrm{c}$ and $0.95\,T_\textrm{c}$ as highlighted by the shaded area in Fig.\,\ref{FMR-alpha}(b). Remarkably, the range of the enhancement in $\alpha_\textrm{eff}$ overlaps with this region.

\begin{figure}[t]
	\centering
	\includegraphics[width=3.4in]{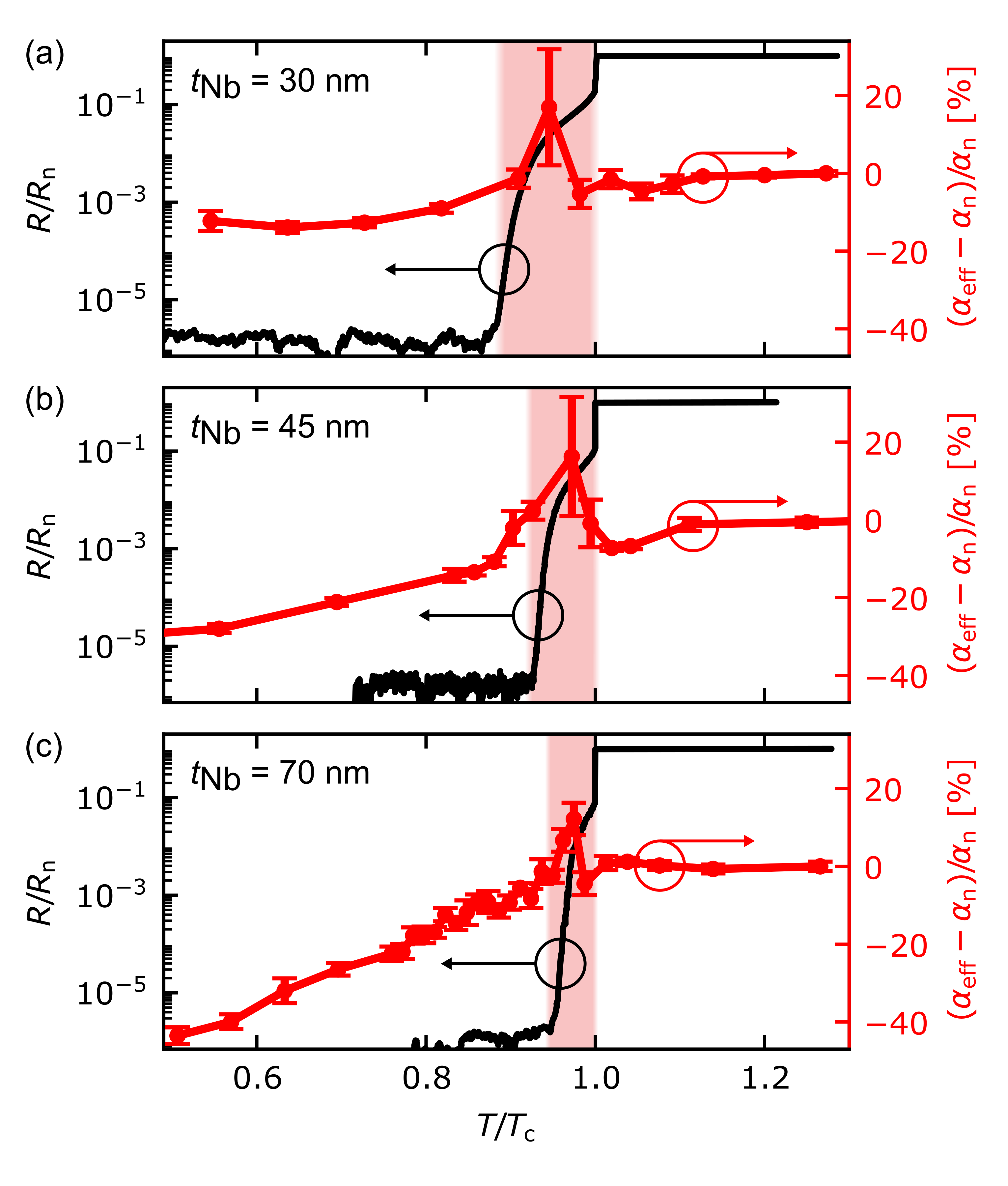}
	\caption{Correlation between excess damping and dissipative transport. The normalized damping $(\alpha_\textrm{eff}-\alpha_\textrm{n})/\alpha_\textrm{n}$ and the normalized resistance $R/R_\textrm{n}$ are shown as a function of $T/T_\textrm{c}$ for (a) $t_\textrm{Nb} = 30$, (b) 45 and (c) \SI{70}{nm}. The resistance is evaluated at $\mu_0 H = \SI{500}{mT}$. The shaded regions mark the intervals in which the transport measurements remain dissipative below $T_\textrm{c}$. The lower limits of these intervals are $0.88\,T_\textrm{c}$ for $t_\textrm{Nb}=\SI{30}{nm}$, $0.92\,T_\textrm{c}$ for $t_\textrm{Nb}=\SI{45}{nm}$, and $0.95\,T_\textrm{c}$ for $t_\textrm{Nb}=\SI{70}{nm}$.}
	\label{DampingThickness}
\end{figure}

To identify the origin of the damping enhancement, we examine its evolution for different Nb thicknesses and compare it directly with the superconducting transport properties.
Figs.\,\ref{DampingThickness}(a-c) summarize the temperature dependence of the normalized damping $(\alpha_\textrm{eff}-\alpha_\textrm{n})/\alpha_\textrm{n}$ for $t_\textrm{Nb}= 30$, 45 and \SI{70}{nm}. All samples exhibit the same characteristic
enhancement of $12$--$17\,\%$ immediately below $T_\textrm{c}$. Upon further cooling, a gradual decrease is observed which is more pronounced for larger $t_\textrm{Nb}$.
By comparing the normalized damping to the normalized resistance $R/R_\textrm{n}$ also shown in Figs.\,\ref{DampingThickness}(a-c), we find that the vortex-creep regime deduced from the resistive transport response matches the temperature range of the enhanced damping for all samples. For $t_\textrm{Nb} = \SI{30}{nm}$, both features occur within the interval $\sim0.88\,T_\textrm{c} < T < T_\textrm{c}$. The lower limit of the interval is replaced by $0.92\,T_\textrm{c}$ for $t_\textrm{Nb} = \SI{45}{nm}$, and by $0.95\,T_\textrm{c}$ for \SI{70}{nm}.

The experimental observations establish three key results. First, the damping enhancement appears exclusively below the superconducting transition. Second, it is confined to the vortex-creep regime. Third, its magnitude follows the dissipative transport properties over a broad range of Nb thicknesses. Together, these observations demonstrate that the additional spin dissipation is intimately linked to mobile vortices rather than superconductivity itself as we will detail below.

Several previously discussed mechanisms can be excluded as the origin of the enhanced damping. Superconducting coherence effects arising from the peaks in the quasiparticle density of states\,\cite{Inoue2017,Yao2018,Mueller2021}, the finite-frequency spin susceptibility in the presence of spin-orbit coupling\,\cite{Silaev2020a}, and Andreev bound states at the S/F interface\,\cite{Silaev2020b,Tanhayi2021} all predict damping enhancements that extend over a temperature range which is several times broader than the narrow maximum of width $\sim0.1\,T_\textrm{c}$ observed here. Magnetic impurity scattering is predicted to enhance the damping near $T_\textrm{c}$ by only $\sim 1\,\%$ which is far below the observed $17\,\%$, and applies explicitly to FM/SC structures without heavy-metal spin sinks\,\cite{Morten2008}.
An apparent enhancement arising from the magnetic-field dependence of the effective transition temperature can likewise be ruled out: as detailed in the Supplemental Material\,\cite{SupplMat}, this artifact would produce an apparent \textit{reduction} of $\alpha_\textrm{eff}$ near $T_\textrm{c}$, opposite to the observation.
Finally, an inhomogeneous, lateral nucleation of superconductivity is difficult to reconcile with the sharp onset of the superconducting regime discussed in the Supplemental Material\,\cite{SupplMat}.

Instead, the systematic correspondence between the damping enhancement and the finite-resistance vortex-creep regime found in Fig.\,\ref{DampingThickness} points to a mechanism associated with mobile vortices. These mobile vortices, driven by an orthogonal gradient in the spin accumulation, provide the additional channel for spin-angular-momentum dissipation. Conversely, once vortex motion becomes suppressed by pinning effects at lower temperatures, the additional damping channel disappears, and the temperature- and Nb-thickness-evolution of $\alpha_\textrm{eff}$ agrees with the gradual freeze-out of quasiparticle states in a spin-singlet condensate\,\cite{Bell2008, Jeon2018, Mueller2021}.

When invoking spin-angular-momentum dissipation through mobile vortices, we have to establish that the vortex motion is indeed driven by spin pumping in our geometry as proposed in Fig.\,\ref{Concept}. The magnetization of the Py layer precesses around its principal direction given by the external field, which is aligned in the sample plane. The precession generates a spin accumulation gradient that is polarized along the same principal direction and decays into the adjacent Nb layers.
As theoretically demonstrated by Refs.\,\cite{Taira2021, Adachi2024}, a spin accumulation gradient can exert a force on vortices which are polarized orthogonally to the gradient.
Hence, in order for the out-of-plane spin accumulation gradient to drive the vortex motion, the vortices must possess an in-plane component. In a strictly parallel field, an in-plane vortex is only stable in a film if $t_\textrm{Nb} \gtrsim \sqrt{5}\,\xi$\,\cite{Tinkham2004}. The coherence length is $\xi = \sqrt{\Phi_0 / 2\pi \mu_0 H_\textrm{c2}}$\,\cite{BardeenStephen1965}, where $\Phi_0$ is the magnetic flux quantum, and the upper critical field $\mu_0 H_\textrm{c2} = \SI{1.15}{T}$ is determined experimentally for the sample with $t_\textrm{Nb} = \SI{30}{nm}$ (see Supplemental Material\,\cite{SupplMat}), yielding $\xi = \SI{16.9}{nm}$. With $\sqrt{5}\,\xi = \SI{37.8}{nm}$, the criterion for the existence of in-plane vortices is met for $t_\textrm{Nb} = 45$ and \SI{70}{nm}.
Notably, as $T$ approaches $T_\textrm{c}$, $\xi$ rises since $\xi(T) \propto 1 / \sqrt{1-T/T_\textrm{c}}$\,\cite{Tinkham2004}. In the decisive temperature window above $0.9\,T_\textrm{c}$, $\xi(T)$ increases to $\SI{53.5}{nm}$. Correspondingly, vortices are expected to be tilted and acquire an out-of-plane component, originating from an unavoidable misalignment of the external magnetic field.
The mechanism discussed above is, hence, effective only on the remaining in-plane component. Therefore, this concept is applicable to all presented samples.

The assignment to mobile vortices is corroborated by the characteristic temperature scales of vortex dynamics in thin-film Nb, where the vortex-glass transition occurs $0.1$--$1$\,K below $T_\textrm{c}$\,\cite{Villegas2005,Altanany2024}, in quantitative agreement with the width of the observed damping maximum. It is further supported by spin Seebeck experiments near $T_\textrm{c}$, which detected a coupling between spin currents and vortex motion in the vortex-liquid phase via the vortex Nernst effect\,\cite{Umeda2018,Sharma2023}, and by the theoretical prediction of an inverse vortex spin Hall effect in type-II superconductors\,\cite{Taira2021,Adachi2024}.
The narrow width of the enhancement follows naturally. The fraction of the film driven into the vortex-creep regime, $\rho_\textrm{f}/\rho_\textrm{n} = H/H_\textrm{c2}(T)$, scales as $(1-T/T_\textrm{c})^{-1}$ and therefore grows by about an order of magnitude between $0.9\,T_\textrm{c}$ and $T_\textrm{c}$, independently of the coherence length and the precise field misalignment. This steep, near-critical growth mirrors the narrow temperature interval over which the damping enhancement is observed.

We emphasize that this areal estimate fixes the temperature scale of the effect but not its sign. A \textit{static} normal fraction absorbs spin no more efficiently than the fully normal film and can therefore only restore the damping towards $\alpha_\textrm{n}$, never exceed it. Surpassing $\alpha_\textrm{n}$ requires vortex \textit{motion}: the dissipation associated with mobile normal cores or, in other words,  viscous flux flow and the time-dependent order parameter of a moving vortex, constitutes a channel with no static, normal-state counterpart. The observed excess damping of $12$--$17\,\%$ is thus consistent with a vortex-motion channel whose efficiency is comparable to that of normal-state spin absorption, rather than requiring an anomalously strong coupling.

Moreover, employing the theoretical framework developed in Ref.\,\cite{Taira2021}, we find that the temperature dependence of the spin conductivity $\sigma^\textrm{(s)}_{xx}$ reproduces the enhancement in $\alpha_\textrm{eff}$ if the in-plane vortex contribution $\delta \sigma^ \textrm{(s)}_{xx}$ is included, see Supplemental Material\,\cite{SupplMat}. $\delta \sigma^ \textrm{(s)}_{xx}$ is governed by the gap in the renormalized masses between the lowest and next-lowest Landau levels, $\widetilde{\alpha}_1 - \widetilde{\alpha}_0 = 2 (H/H_\textrm{c2})r_\textrm{gap}$, where the renormalization factor $r_\textrm{gap} = 0.02$ was used to fit the width and height of the enhancement to the experimental observation.
The low value of $r_\textrm{gap}$ likely points towards the strong renormalization of Landau levels from the vortex-liquid into the vortex-creep regime. The match of theoretical prediction and experimental observation underpins our interpretation of the enhancement in $\alpha_\textrm{eff}$ through spin-angular-momentum dissipation through mobile vortices.

Our results identify mobile vortices as a previously unexplored channel for spin-angular-momentum dissipation in superconducting spintronic heterostructures. Beyond providing new insight into dynamic spin transport in superconductors, these findings establish ferromagnetic resonance as a sensitive probe of dissipative vortex dynamics and open new opportunities for controlling spin relaxation through the superconducting vortex state.

\begin{acknowledgments}
The authors thank Christoph Strunk, Dhavala Suri and Marco Aprili for insightful discussions. This work was funded by Deutsche Forschungsgemeinschaft (DFG, German Research Foundation) through Project-ID 314695032 - SFB 1277 (Subproject A08), and by the JSPS KAKENHI Grant (No. 26K00658).\\
\end{acknowledgments}

\textit{Author contribution statement} --  CHB and LC conceptualized the experiments. MM, AB and FW performed the measurements. TNGM provided samples. MM, AB and LC analyzed the data. HA performed the calculations. MM, LC, HA and CHB wrote the manuscript with inputs from all authors. All authors contributed to the discussions.\\

\textit{Data availability} -- The data that support the findings of this article are openly available%\,\cite{Zenodo}.

\bibliography{references.bib}

\clearpage
\onecolumngrid
\section*{Supplemental Material}
\setcounter{figure}{0}
\setcounter{table}{0}

\renewcommand{\thefigure}{S\arabic{figure}}
\renewcommand{\theequation}{S\arabic{equation}}

\numsection{\\Details on the FMR measurement and evaluation method}
\label{SI:sec:meas}

\begin{figure}[tbh]
    \centering
    \includegraphics[width=5in]{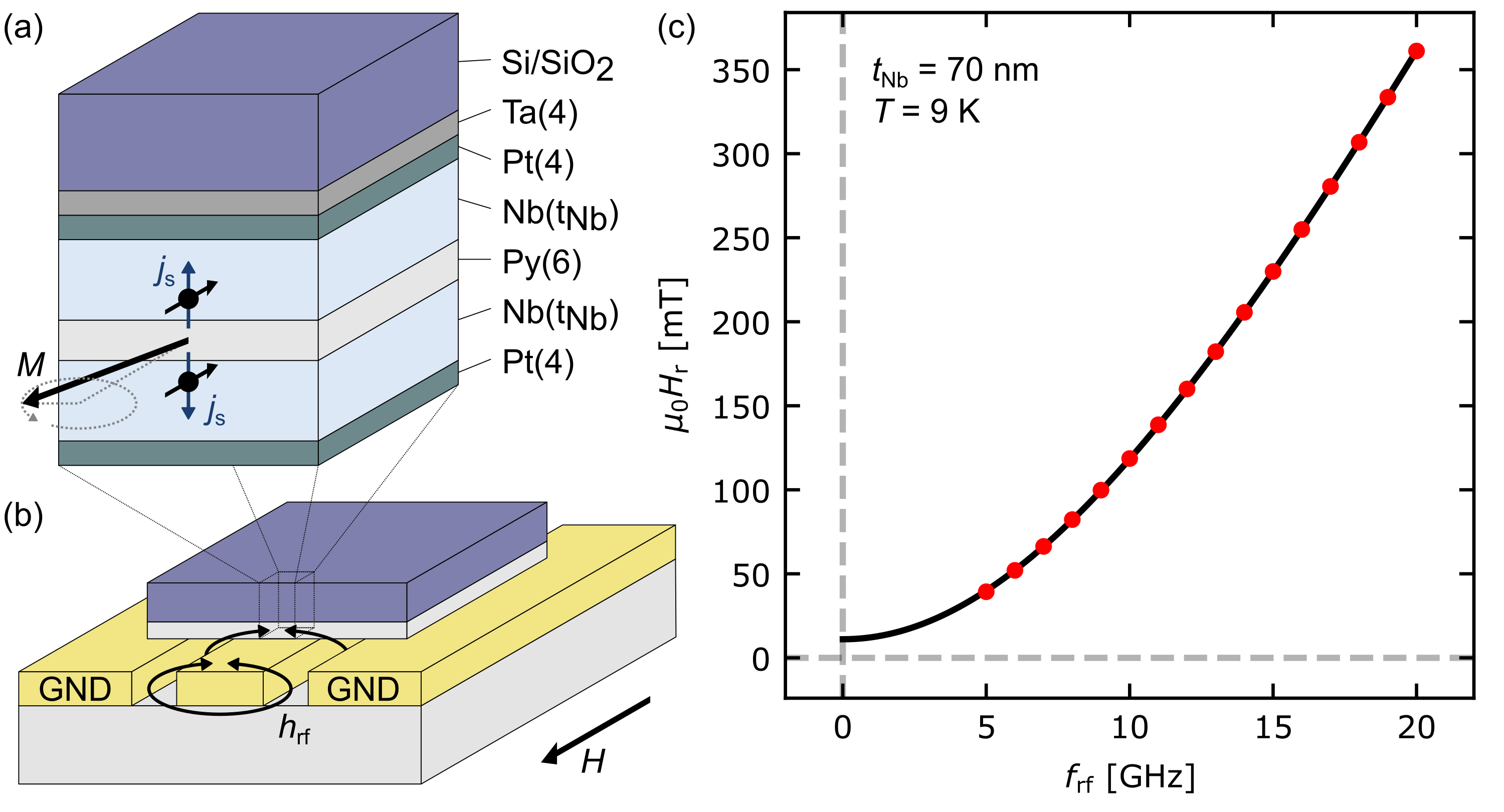}
    \caption{Principle of the FMR measurements. (a) Stack sequence where the individual layer thicknesses are given in nm. $M$ denotes the precessing magnetization of the Py layer, and $j_\textrm{s}$ is the spin current pumped into the Nb layers. (b) Schematic measurement geometry described in the text. The CPW is made of \SI{40}{\micro m} thick Cu. The signal line is \SI{340}{\micro m} wide, the gap measures \SI{135}{\micro m}, and the ground lines extend beyond the sample borders. (c) Example of the frequency dependence of $H_\textrm{r}$. The black line represents a fit according to the modified Kittel Eq.\,(\ref{Kittel}).}
    \label{FMRmethod}
\end{figure}

The precession of the Py magnetization $M$ is excited using a full-film ferromagnetic resonance (FMR) technique where the sample, schematically shown in Fig.\,\ref{FMRmethod}(a), is mounted face-down on a coplanar waveguide (CPW), as sketched in Fig.\,\ref{FMRmethod}(b). A $\sim\SI{1}{\micro m}$ thick layer of photoresist on the sample's surface ensures electric insulation from the CPW. An external field $H$ saturates $M$ along the CPW. A radio-frequency (rf) current is applied to the signal line of the CPW, generating an alternating magnetic field $h_\textrm{rf}$. The power and frequency of the rf signal are kept constant while $H$ is swept through the resonance condition. The absorbed rf signal is rectified by a Schottky diode and detected by a lock-in amplifier that is locked to a modulation of $H$ of less than \SI{1}{mT} at a frequency of 13 -- \SI{1311}{Hz}.

Fig.\,2(a) of the main text depicts a typical resonance curve acquired using the sample with $t_\textrm{Nb}=\SI{70}{nm}$ at a frequency of $f_\textrm{rf}=\SI{10}{GHz}$ at \SI{9}{K}, which is above $T_\textrm{c}=\SI{7.9}{K}$ of this sample (see Supplemental Note~2). The resonance curves are fitted by the sum of a symmetric and an antisymmetric Lorentz function:
\begin{align} \label{Lorentz}
    V_\textrm{FMR} = V_0 + A \cdot \frac{ -2(H - H_\textrm{r}) \cdot \Delta H \cos\varphi + \left( (\Delta H)^2 - (H - H_\textrm{r})^2 \right) \sin\varphi }{ \left( (H - H_\textrm{r})^2 + (\Delta H)^2 \right)^2 }
\end{align}
where $H_\textrm{r}$ denotes the resonance field and $\Delta H$ the half width at half maximum of the resonance. $A$ is the amplitude and $\varphi$ is a phase factor to balance the symmetric and antisymmetric contributions.

Resonance curves are measured at various excitation frequencies between 5 and \SI{20}{GHz} to extract $H_\textrm{r}$ and $\Delta H$ as a function of $f_\textrm{rf}$. Figure\,\ref{FMRmethod}(c) shows $H_\textrm{r}(f_\textrm{rf})$ at \SI{9.0}{K}. The curve is fitted by the modified Kittel formula\,\cite{Kittel1948}
\begin{align} \label{Kittel}
    H_\textrm{r} = \frac{1}{2} \cdot \biggl(  -\bigl( M_\textrm{eff} + 2H_\textrm{an} \bigr) \,+ \,\sqrt{ \bigl( M_\textrm{eff} \bigr)^2 + \bigl( 4\pi f_\textrm{rf}/(\gamma\,\mu_0) \bigr)^2 } \biggr)
\end{align}
where $\gamma$ is the gyromagnetic ratio, $M_\textrm{eff}$ denotes the effective magnetization of the Py layer and $H_\textrm{an}$ is an anisotropy term.
The extrapolated zero-frequency intercept can be interpreted directly as $-H_\textrm{an}$. At the superconducting transition, a significant increase of $H_\textrm{an}$ is observed, as discussed in Supplemental Note~2. The evaluation of $\Delta H$ and the extraction of the effective Gilbert damping $\alpha_\textrm{eff}$ is discussed in the main text. 
This method is applied to all samples with $t_\textrm{Nb} = 30$, 45 and \SI{70}{nm} in a temperature range between \SI{2}{K} and \SI{10}{K}.

\newpage

\numsection{\\Extraction of $T_\textrm{c}$ from FMR measurements}\label{SI:sec:HanSC}

\begin{figure}[tbh]
    \centering
    \includegraphics[width=3.4in]{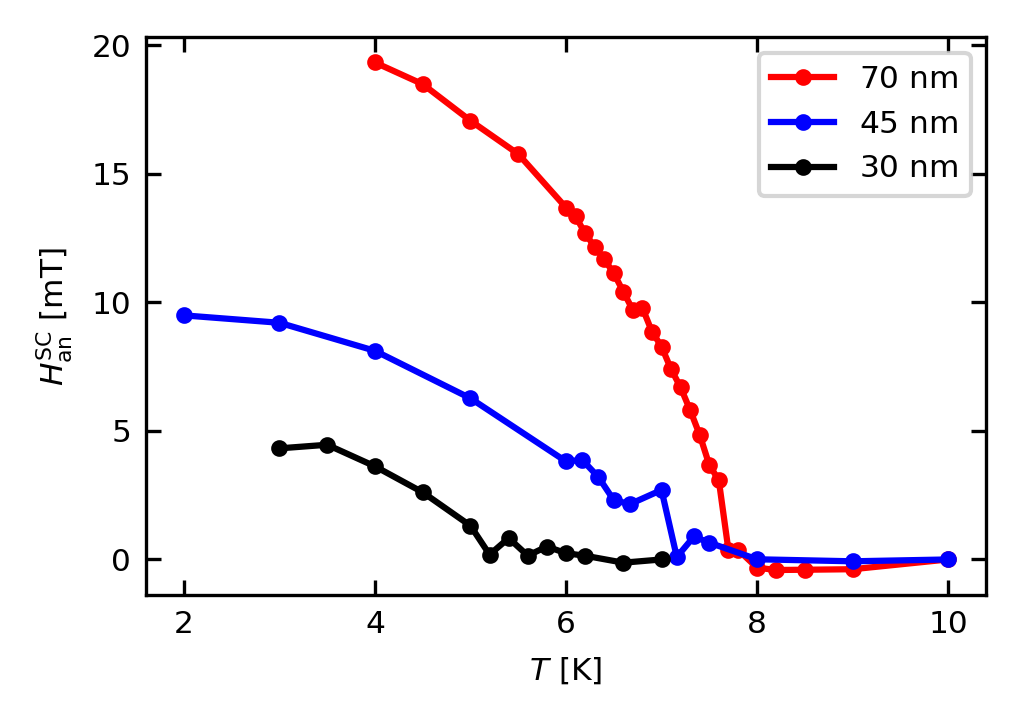}
    \vspace{-20pt}
    \caption{Temperature-dependence of the superconductivity-induced anisotropy field $\mu_0 H_\textrm{an}^\textrm{SC} = \mu_0 H_\textrm{an} - \mu_0 H_\textrm{an}|_{T > T_\textrm{c}}$ for $t_\textrm{Nb}=30$, 45 and \SI{70}{nm}. The transition temperature $T_\textrm{c}$ for the respective samples are extracted from the onset of the increase in $\mu_0 H_\textrm{an}^\textrm{SC}$. The resulting values are $T_\textrm{c}=\SI{5.5}{K}$ for $t_\textrm{Nb}=\SI{30}{nm}$, \SI{7.2}{K} for $t_\textrm{Nb}=\SI{45}{nm}$ and \SI{7.9}{K} for $t_\textrm{Nb}=\SI{70}{nm}$.}
    \label{FluxFocusing}
\end{figure}

In the following, the anisotropy term $H_\textrm{an}$ is analyzed. For the purpose of this report, we focus on the superconductivity-induced effective anisotropy field $H_\textrm{an}^\textrm{SC} = H_\textrm{an} - H_\textrm{an}|_{T > T_\textrm{c}}$, which is obtained by subtracting the normal-state anisotropy field $H_\textrm{an}|_{T > T_\textrm{c}}$. This offset varies between the samples due to the random orientation between the anisotropy axis and the external magnetic field.
The temperature-dependence of $H_\textrm{an}^\textrm{SC}$ is shown in Fig.\,\ref{FluxFocusing} for all samples. When cooling down from temperatures above $T_\textrm{c}$, $H_\textrm{an}^\textrm{SC}$ changes abruptly and increases up to 20 (10, 5) mT for $t_\textrm{Nb} = 70$ (45, 30) nm. The onset of the increase is used to determine $T_\textrm{c} = 7.9$ (7.2, 5.5) K for $t_\textrm{Nb} = 70$ (45, 30) nm. The emergence of an anomalous anisotropy term in the superconducting regime has been reported before\,\cite{Li2018, Jeon2019, Golovchanskiy2020, Tian2025}, but the physical origin remains debated. For example, Ref.\,\cite{Silaev2022} suggests a magnon dispersion altered by the Anderson-Higgs mechanism through the superconducting proximity as the source of $H_\textrm{an}^\textrm{SC}$. An alternative explanation is based on flux focusing by the Meissner effect in the Nb layers which reinforces the field perceived by the Py magnetization in addition to the externally applied field $H$\,\cite{Jeon2019}. The increase in $H_\textrm{an}^\textrm{SC}$ with increasing $t_\textrm{Nb}$ observed in Fig.\,\ref{FluxFocusing} is consistent with this picture\,\cite{Tian2025}: Since all samples are thinner than the London penetration depth of thin-film Nb, $\lambda_\textrm{L}>\SI{100}{nm}$\,\cite{Gubin2005}, the total flux expelled from the superconducting volume still grows with $t_\textrm{Nb}$.

\newpage

\numsection{\\Details on the transport measurements}
\label{SI:sec:transport}

\begin{figure}[tbh]
\centering
\includegraphics[width=\linewidth]{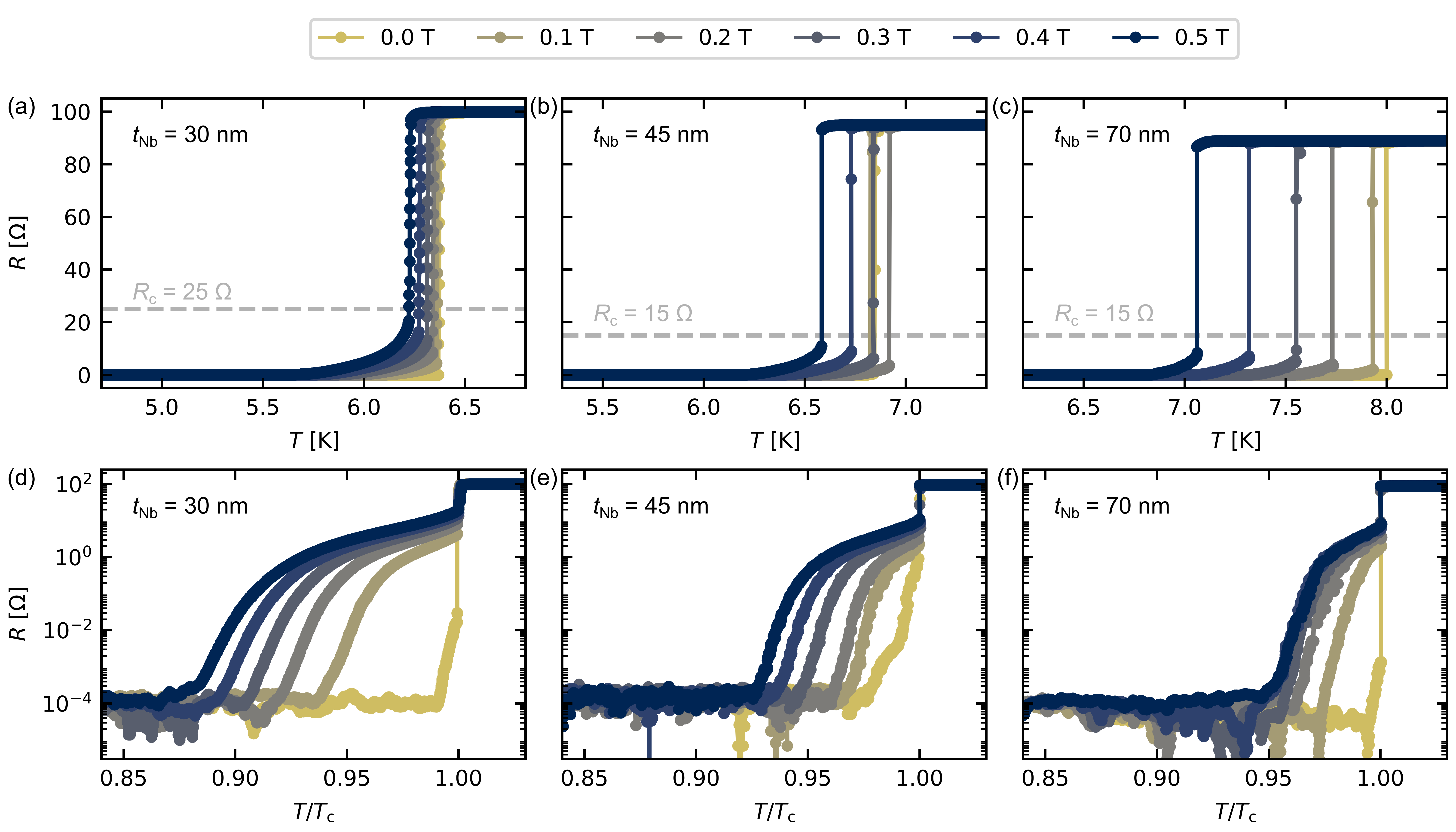}
\caption{Four-terminal resistance measurements under in-plane magnetic field $H$. (a-c) $R(T)$ at various values of $H$ for the samples with $t_\textrm{Nb} = 30$, 45 and \SI{70}{nm}. The horizontal dashed lines indicate the threshold resistance $R_\textrm{c} = 25$, 15 and \SI{15}{\Omega} for the respective samples from which the critical temperature $T_\textrm{c}$ is determined. (d-f) Plotting the same $R(T)$-curves against the normalized temperature $T/T_\textrm{c}$ on a semi-logarithmic scale reveals the low-resistance tail attributed to the vortex-creep regime.}
\label{SI:fig:transport}
\end{figure}

For the transport measurements, the samples are patterned by optical lithography and Ar sputtering into \SI{1.7}{mm} long stripes. The width of the stripe is $w = 20.0$ (13.33, 8.57)\,\SI{}{\micro m} for the sample with $t_\textrm{Nb} = 30$ (45, 70)\,nm such that cross-section $wt_\textrm{Nb} = \SI{0.6}{\micro m^2}$ is constant for all samples, thus producing the same current density at a fixed current. A driving current of \SI{1}{mA} is applied and the voltage across the stripe is measured in a four-point geometry as a function of temperature. The magnetic field $H$ is applied along the stripe. By performing the measurement while cooling down across the superconducting transition at the respective $H$, the sample is expected to be in a saturated vortex state.
The resulting $R(T)$-curves are depicted in Figs.\,\ref{SI:fig:transport}(a-c).
The normal-state resistance is $R_\textrm{n} \approx 90-\SI{100}{\ohm}$ for all samples due to the constant $wt_\textrm{Nb}$-product.
The transition into the superconducting state is characterized by a sharp drop in resistance at the effective critical temperature $T_\textrm{c}(H)$, followed by an extended low-resistance tail. $T_\textrm{c}(H)$ is reduced by 0.5--\SI{1}{K} for $H = \SI{0.5}{T}$ compared to the field-free $T_\textrm{c}$. 
To eliminate the $H$-dependence, the normalized temperature is defined as $T/T_\textrm{c}(H)$, where $T_\textrm{c}(H)$ is extracted at the threshold resistance $R_\textrm{c}$ as indicated in Figs.\,\ref{SI:fig:transport}(a-c). This is reasonable when comparing transport and FMR measurements since the values for $T_\textrm{c}$ extracted from $H_\textrm{an}^\textrm{SC}$ (see Supplemental Note~2) are inherently affected by $H$.
To investigate the low-resistance feature, the data is reproduced  in Figs.\,\ref{SI:fig:transport}(d-f) on a semi-logarithmic scale and with a normalized axis $T/T_\textrm{c}(H)$. With increasing $H$, the width of the tail increases for all samples, consistent with thermally activated vortex creep\,\cite{Eley2021}. At $H=\SI{0.5}{T}$, the vortex-creep regime extends to 0.88 (0.92, 0.95)\,$T_\textrm{c}$ for the sample with $t_\textrm{Nb} = 30$ (45, 70)\,nm. The relevance of the value $H=\SI{0.5}{T}$ derives from the fact that the field sweeps applied during the FMR measurements cover a magnetic field range of up to \SI{0.5}{T}. Since the vortex state is highly hysteretic, it is justified to use this field value for the discussion of the vortex-creep regime in the FMR experiments.

\newpage

\numsection{\\Excluding the $H$ dependence of $T_\textrm{c}$ as the source of the enhancement in $\alpha_\textrm{eff}$}\label{SI:sec:TcH}

\begin{figure}[tbh]
    \centering
    \includegraphics[width=3.4in]{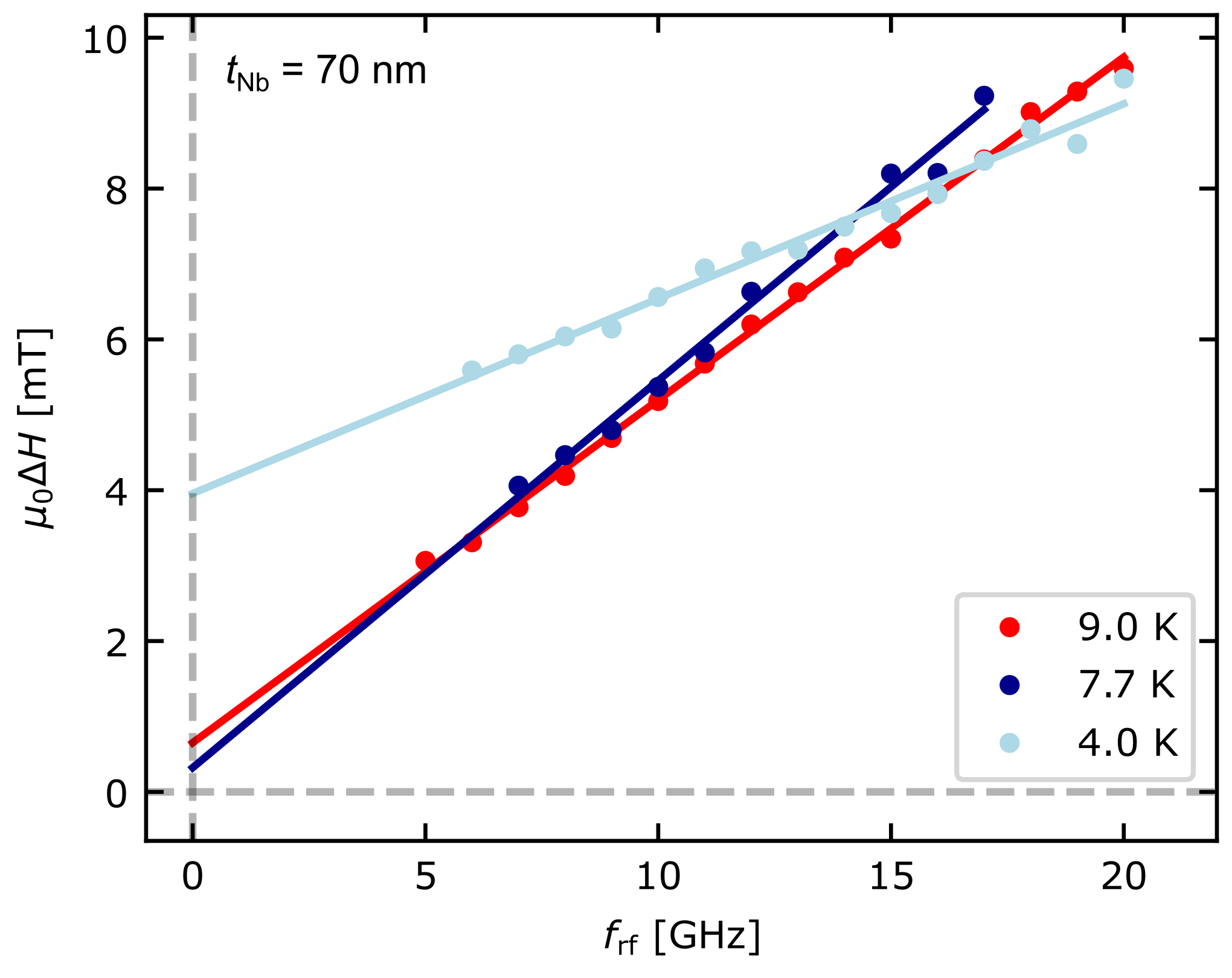}
    \caption{Selection of frequency-dependent $\Delta H$ curves at different temperatures with the respective linear fits.}
    \label{DeltaH-frf}
\end{figure}

Here, we elaborate on the possibility of the enhancement of $\alpha_\textrm{eff}$ near $T_\textrm{c}$ being caused by the shift of the effective $T_\textrm{c}$ with $H$. Transport measurements (see Supplemental Note~3) reveal that the effective $T_\textrm{c}(H)$ is decreased by 0.5--\SI{1}{K} for $H=\SI{0.5}{T}$ with respect to the field-free case. Since the values of $H_\textrm{r}$ increase with $f_\textrm{rf}$, resonances at larger $f_\textrm{rf}$, in principle, experience a lower effective $T_\textrm{c}$. When measuring close to the superconducting transition, the superconducting state may thus only be accessible by lower frequencies while higher ones still probe the normal state. However, as shown in Fig.\,\ref{DeltaH-frf}, changes in $\Delta H$ across the superconducting transition are, for a given $f_\textrm{rf}$, dominated by the increasing zero-frequency intercept $\Delta H_0$ rather than the slope that is proportional to $\alpha_\textrm{eff}$. Since $\Delta H_0$ is strongly increased in the superconducting regime (see light-blue line at \SI{4.0}{K}), this effect would produce an apparent reduction in $\alpha_\textrm{eff}$, which is not observed. Instead, the enhancement of $\alpha_\textrm{eff}$ causes a minor apparent decrease in the zero-frequency intercept.

\newpage

\numsection{\\Experimental determination of the coherence length}
\label{SI:sec:meas}

\begin{figure}[tbh]
\centering
\includegraphics[width=3in]{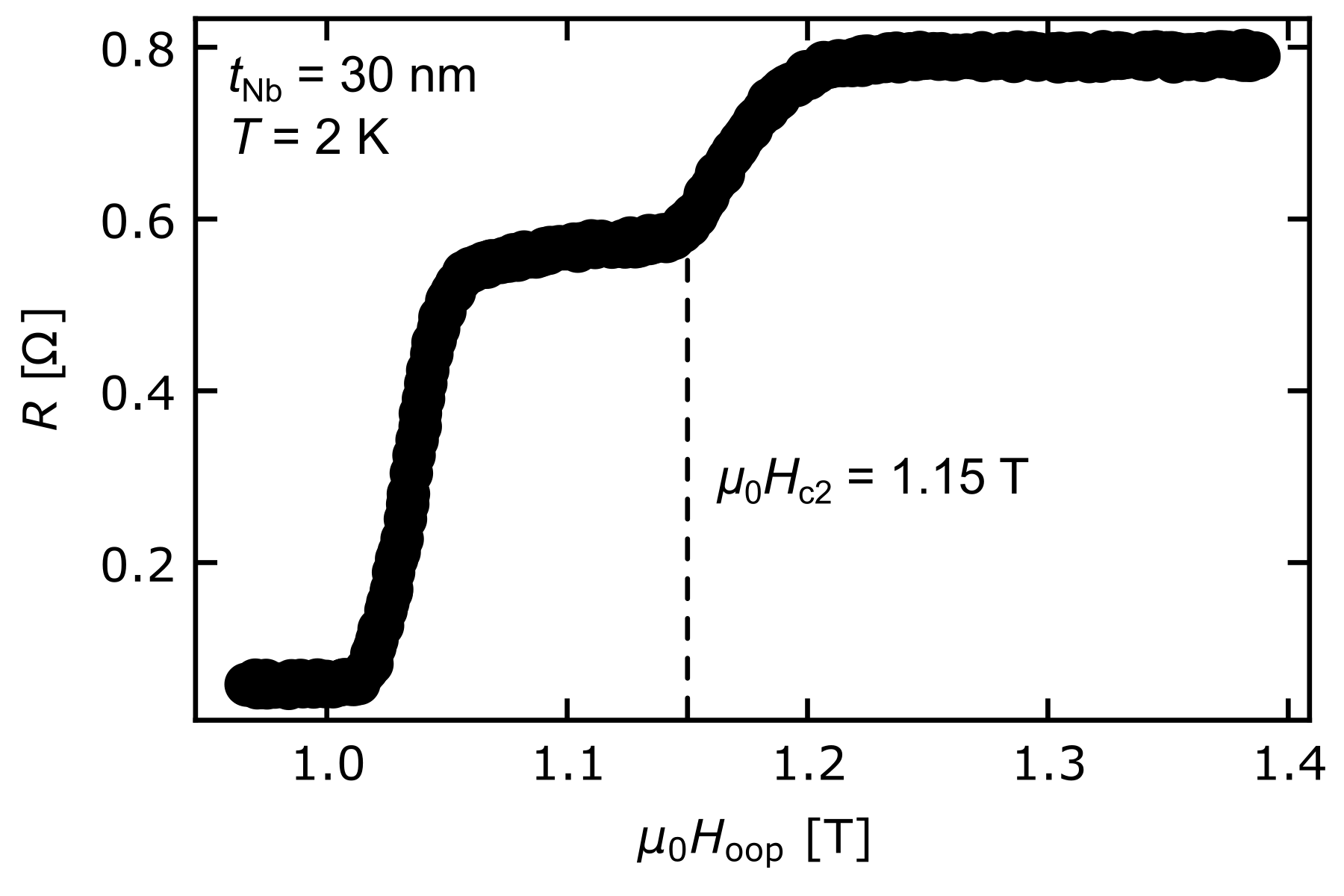}
\caption{Four-terminal resistance measurements under out-of-plane magnetic field $H_\textrm{oop}$ for $t_\textrm{Nb} = \SI{30}{nm}$. The data is acquired on an unpatterned full-film sample, resulting in different absolute resistance values in the normal and vortex-creep regime compared to Fig.\,\ref{SI:fig:transport}.}
\label{SI:fig:Boop}
\end{figure}

The coherence length $\xi$ is extracted from the upper critical field $\mu_0 H_\textrm{c2}$ measured under an out-of-plane magnetic field $H_\textrm{oop}$. $\mu_0 H_\textrm{c2} = \SI{1.15}{T}$ is determined for the sample with $t_\textrm{Nb} = \SI{30}{nm}$ in Fig.\,\ref{SI:fig:Boop}, yielding $\xi = \sqrt{\Phi_0 / 2\pi \mu_0 H_\textrm{c2}}= \SI{16.9}{nm}$. Between $\mu_0 H_\textrm{oop} = \SI{1}{T}$ and $\mu_0 H_\textrm{c2}$, the finite resistance of the vortex-creep regime is observed.

\newpage

\numsection{\\Theoretical model for the spin conductivity in the presence of mobile vortices}
\label{SI:sec:theory}

In this part, assuming that the in-plane moving vortices play the dominant role in the spin dissipation, we calculate the spin conductivity near the superconducting transition into the vortex state.

We consider exactly the same situation as in Ref.\,\cite{Taira2021}, and, following the same notation, calculate the (longitudinal) spin conductivity using the simplest approximation of dealing with fluctuation effects (Hartree approximation\,\cite{Ullah1991}). We start from the Kubo formula 
%%%
\begin{equation}
  \sigma^{\rm (s)}_{xx} = \left. \frac{1}{\ui \Omega} \Big( K^{{\rm (s)} R}_{xx} (\Omega)- K^{{\rm (s)} R}_{xx} (0)\Big) \right|_{\Omega \to 0}, 
\end{equation}
%%%
where $K^{{\rm (s)} R}_{xx} (\Omega)$ is the spin current retarded correlation function. The spin conductivity can be given as a sum of normal state contribution $\sigma^{\rm (s)}_N$ and the vortex contribution $\delta \sigma^{\rm (s)}_{xx}$ as
%%%
\begin{equation}
  \sigma^{\rm (s)}_{xx} = \sigma^{\rm (s)}_N + \delta \sigma^{\rm (s)}_{xx},
  \label{eq:sigmaS01}
\end{equation}
%%%
where $\sigma_N^{\rm (s)}= 1/\rho_N$ with the normal state electrical resistivity $\rho_N$. 

Proceeding with a calculation similar to Ref.\,\cite{Taira2021}, the second term of the above equation is obtained from the retarded correlation function 
%%%
\begin{eqnarray}
  \delta K^{{\rm (s)} R}_{xx} &=& \frac{\ui \Omega}{4 \pi T \ell_H^4} 
  \sum_N (N+1) \int_\omega \int_{q} \frac{(t^{'{\rm (s)}})^2 \omega^2}{\sinh^2 \left( \frac{\omega}{2T} \right)} \nonumber \\
      &\times& {\rm Re} \Big[ L^R_{N,q}(\omega) L^A_{N+1,q}(\omega) \Big],  
\end{eqnarray}
%%%
where we introduced a shorthand notation $\int_\omega= \int_{-\infty}^\infty \frac{d \omega}{2 \pi}$ and $\int_q= \int_{-\infty}^\infty \frac{d q}{2 \pi}$. Here, $\ell_H= 1/\sqrt{2|e|H}$ is the magnetic length, $t^{'{\rm (s)}}= -4 |e| \xi_0^2 {\mu_s}/{8 T^2}$, and
%%%
\begin{equation}
  L^R_{N,q}(\omega)= \frac{1}{\alpha_{N,q} - \ui \omega/\gamma }
\end{equation}
%%%
is the retarded fluctuation propagator. In this equation, $\alpha_{N,q}= \alpha_N + \xi_0^2 q^2$ with the mass of the $N$th Landau level, 
%%%
\begin{equation}
  \alpha_{N}= \frac{T- T_{c0}}{T_{c0}} + h (2N+1), 
\end{equation}
%%%
where $h= H/H_{c2}(0)$, $T_{c0}$ is the superconducting transition temperature at zero magnetic field, and $\gamma$ is the dissipation coefficient ($=8 T_{c0}/\pi$ in the BCS limit). The advanced correlator is given by $L^A_{N,q}(\omega)= (L^R_{N,q}(\omega))^*$. 

As in Ref.\,\cite{Taira2021}, we use the approximation that the lowest Landau level ($N=0$) fluctuations dominate the vortex liquid region, and consider the mass renormalization of the lowest Landau level by the Hartree approximation,
%%%
\begin{equation}
  \widetilde{\alpha}_0 = \alpha_0 + \frac{h}{2 \pi} \frac{b'}{\sqrt{\widetilde{\alpha}_0}},   
\end{equation}
%%%
where $b'$ is the dimensionless quartic term of the Ginzburg-Landau coefficient, which is related to the specific heat jump $\Delta C$ under zero magnetic field as $b'= 1/\Delta C \xi_0^3$. 
%%%

Now, by performing the $\omega$ integral as well as the $q$ integral, we obtain 
%%%
\begin{eqnarray}
  \delta \sigma^{{\rm (s)}}_{xx} &=& \frac{\mu_s^2 \gamma}{32\sqrt{2} R_{\cal Q} \xi_0 T^3} 
  \frac{1}{\sqrt{\widetilde{\alpha}_0 + \widetilde{\alpha}_1}}, 
\end{eqnarray}
%%%
where $R_{\cal Q}= \pi/2 |e|^2$ is the quantum resistance, and $\widetilde{\alpha}_1$ is the renormalized mass of the next lowest Landau level. For this renormalized mass $\widetilde{\alpha}_1$, while an approximation $\widetilde{\alpha}_1 \approx \widetilde{\alpha}_0+ 2h$ was used in Ref.\,\cite{Ikeda1991} (see Eq.(2.5) therein), we use $\widetilde{\alpha}_1 \approx \widetilde{\alpha}_0+ 2h \, r_{\rm gap}$ by introducing a new parameter $r_{\rm gap}$. 

\begin{figure} [t] 
  \begin{center}
    \includegraphics[width=10cm]{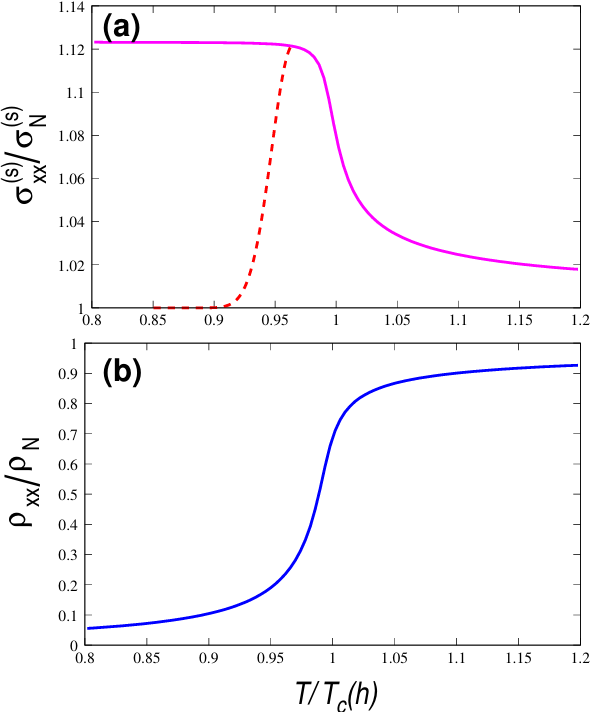}
  \end{center}
  \caption{(a) Temperature dependence of $\sigma^{\rm (s)}_{xx}$ calculated using Eq.\,(\ref{eq:sigmaS01}) for a parameter choice of $b'=0.005$, $\rho_N/R_{\cal Q} \xi_0= 0.15$, $\mu_s/T_{c0}= 2.0$, $\gamma/T_{c0}= 8.0$, $r_{\rm gap}= 0.02$, and $h=0.3$. Here, $T_c(h)$ means the transition temperature in the mean-field approximation at magnetic field $h=H/H_{c2}(0)$. (b) Temperature dependence of the resistivity with the same parameters as (a).}
  \label{fig:sigma}
  \end{figure}

Figure\,\ref{fig:sigma}(a) shows an example of temperature dependence of the spin conductivity, calculated from Eq.\,(\ref{eq:sigmaS01}) for a parameter choice of $b'=0.005$, $\rho_N/R_{\cal Q} \xi_0= 0.15$, $\mu_s/T_{c0}= 2.0$, $\gamma/T_{c0}= 8.0$, $r_{\rm gap}= 0.02$, and $h=0.3$. The solid line is the calculated result for $\sigma^{\rm (s)}_{xx}$, and the dotted line represents an expected behavior when we take into account the vortex pinning effects.

Figure\,\ref{fig:sigma}(b) shows an example of temperature dependence of the electrical resistivity, using the same parameters as (a).

%%%%%%%%%%%%%%%%%%%%%%%%%%%%%%%%%%%%% 
%%%%%%%%%%%%%%%%%%%%%%%%%%%%%%%%%%%%%

\end{document}